# Far-Field Tunable Nano-focusing Based on Metallic Slits Surrounded with Nonlinear-Variant Widths and Linear-Variant Depths of Circular Dielectric Grating

Peng-Fei Cao[1], Ling Cheng[1,2], Xiao-Ping Zhang[1*], Wei-Ping Lu[2], Wei-Jie Kong[1], Xue-Wu Liang[1]

[1] School of Information Science and Engineering, Lanzhou University
Lanzhou 730000，China

[2] Department of Physics, School of Engineering and Physical Sciences, Heriot-Watt University
Edinburgh, EH14 4AS, UK

**ABSTRACT** – In this work, we design a new tunable nanofocusing lens by the linear-variant depths and nonlinear-variant widths of circular grating for far field practical applications. The constructively interference of cylindrical surface plasmon launched by the subwavelength metallic structure can form a subdiffraction-limited focus, and the focal length of the this structures can be adjusted if the each groove depth and width of circular grating are arranged in traced profile. According to the numerical calculation, the range of focusing points shift is much more than other plasmonic lens, and the relative phase of emitting light scattered by surface plasmon coupling circular grating can be modulated by the nonlinear-variant width and linear-variant depth. The simulation result indicates that the different relative phase of emitting light lead to variant focal length. We firstly show a unique phenomenon for the linear-variant depths and nonlinear-variant widths of circular grating that the positive change and negative change of the depths and widths of grooves can result in different of variation trend between relative phases and focal lengths. These results paved the road for utilizing the plasmonic lens in high-density optical storage, nanolithography, superresolution optical microscopic imaging, optical trapping, and sensing.

1. INTRODUCTION

The resolution of almost all conventional optical system is indispensably governed by the diffraction limit. This resolution limit can be overcome by use of focusing the evanescent waves in the near-field region. The concept of "superlens"[1], which can provide higher resolution beyond the diffraction limit have been proved[2-5], was proposed firstly by Pendry in 2000. Plasmonic lens is attracting much interest recently due to its unique feature of exordinary enhanced transmission and super-resolution[6-7]. Plasmonic lens always consisted by metal and dielectric and can excite Surface Plasmon Polarisons (SPPs)[8-10]. It means that we can focus the evanescent components of an illuminated object in the near-field region with subdiffraction-limit resolution[11]. This allows them to break the conventional barrier of diffraction limit, and leads to the formation of concentrated sub-wavelength light spots on the order of nanometers.

---

[*] Corresponding author：zxp@lzu.edu.cn

Researches on manipulating SPPs are mostly focused on specific spatial distribution of subwavelength metallic structures. Haofei Shi et al.[12] propose a novel structure to manipulate beam, where a metallic film is perforated with a great number of nano-slits with specifically designed widths and transmitted light from slits is modulated and converged in free space. The slits transport electromagnetic energy in the form of SPPs in nanometric waveguides and provide desired phase retardations of beam manipulating with variant phase propagation constant. Nonetheless, no one can change focusing position of plasmonic lens by using this method. Some research groups show focusing position of plasmonic lens can be shifted by employing a group one-dimensional grating of groove depths in curved distribution[13-15]. Interestingly, it has been numerically found that the relative phase at the exit end of the slit increases steadily with the increasing groove depth, making it possible to achieve continuous phase retardation by simply designing surrounding grooves with stepped depths. However, the variable range of focusing position is less than 0.8[$\mu$m], and it is difficult to achieve far-field nanofocusing.

The SPPs can be coupled into the radiating form and let higher spatial component contribute to superposition at the far-field focal region to produce subwavelength focal spot with desired dimensions. Some novel two dimensional structures can be researched, such as annular rings[16], curved chains of nanoparticles or nanoholes[17], plasmonic microzone plate (PMZP)-like or chirped slits[18, 19]. Ref.[20] took advantage of tapered dielectric grating to couple SPPs launched by two subwavelength metallic slits, therefore it can obtain the far-field focus by improving the irradiative field. Zhang et al.[21] present a simple plasmonic lens composed of an annular slit and a single concentric groove. The subwavelength groove can scatter the SPPs and constructively interfere to get a far-field focal spot. The focal length can be adjusted by changing the groove diameter. However, the intensity at the focal spot is very weak because the scattering loss results in low focusing efficiency. In our previous work, we present a simple method to realize far-field nanofocusing by utilizing dielectric surface grating upon the circular plasmonic lens[22]. In this scheme, a blocking chip is used to optimize the far-field plasmonic lens in order to depress the energy focusing at the incident space, and then the focusing efficiency is enhanced further. However, the focusing position shift also is kept within narrow limits, and reason why focal length can be modulated is not discussed.

In this paper, we present a novel method to realize far-field tunable nanofocusing by the linear-variant depths and nonlinear-variant widths of circular grating based on our previous work[22]. The larger spectrum of focusing position shift ($3.3\lambda \sim 6\lambda$) is obtained and the reason for adjusting focal length is analyzed. The circular grating can modulate the cylindrical surface plasmon waves launched by annular metallic slits, and then part of the evanescent components can be coupled into propagation waves, which can be eradiated to the far field region and constructively interfered to cause the superfocusing effect. The full wave simulations illustrate that linear-variant depths and nonlinear-variants width of grooves play an important role for varying focus position. The relation between varying focal length and phases can be obtained, and we firstly discover that the positive change and negative change of the depths and widths of circular grating can result in different of variation trend between relative phases and focal lengths. In positive change case, the rule of focusing points shift is the same as phase changing rule. On the contrary, the rule of focusing points shift has an inverse relationship to phase changing rule in negative change case. Our far-field tunable nanofocusing scheme can supply for the requirement of practical applications such as nanolithography, superresolution optical microscopic imaging, optical trapping, and sensing.

**2. SPW DISPERSION RELATION AND WORK MODES ANALYSES**

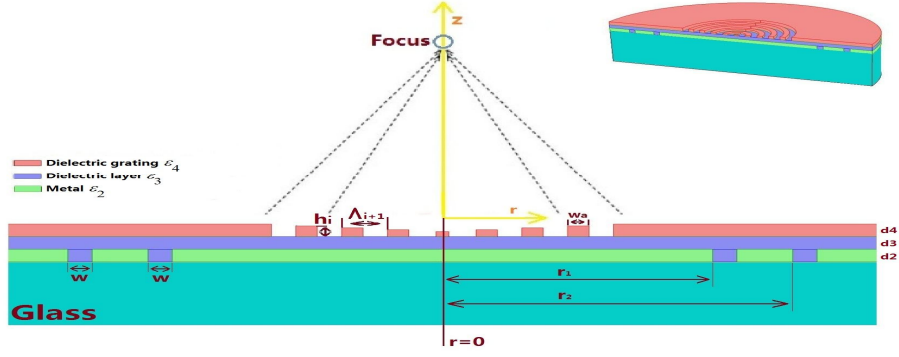

**Fig. 1**. Schematic of the far-field nanofocusing plasmonic lens, as well as the half of the three-dimensional model is depicted at the *top right corner*.

The cross section of our far-field tunable nanofocusing lens is depicted in the r–z plane, as shown in Fig. 1. The lower part is an ordinary circular plasmonic lens with several slit rings milled into metal film upon the glass substrate. Here high-intrinsic-loss metal such as titanium (Ti) or chrome (Cr) is selected, as it is benefit for the subwavelength annular slits exciting diffracted cylindrical waves to launch the cylindrical surface plasmon (CSP) predominantly[18]. Multiple annular slits are milled into the metallic film; the radius and the width of each slit are denoted as $r_j$ and $w$, respectively. A dielectric layer with high refractive index ($n_2$) is adjacent to the metal film, smoothing the metallic surface and protecting the metallic thin film from sulfuration or oxidation as well. A circular dielectric grating is added topside, acting as a modulation device. In order to modulate the focal length, the depth ($h_i$) and width ($\Lambda_i$) of circular grating can be varied. Cross-section profile of circular grating in radial direction can be depicted by Fig. 2, and linear-variant depth and nonlinear-variant width can be expressed as follows,

$$h_i = h_1 - h_d \times (i-1) \tag{1}$$

$$\Lambda_i = \begin{cases} w_a + h_{max}/2, i = max \text{ denote the number of the midpoint of the grating} \\ w_a + (h_{max-1} - h_{max}/2), i = max - 1 \\ w_a + h_{i+1}, \quad i < max - 1 \end{cases} \tag{2}$$

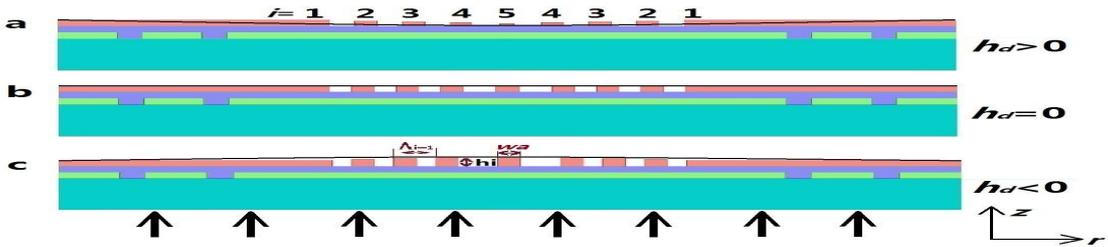

**Fig. 2**. Schematic view of the traced linear-variant depth and nonlinear-variant width profile: $w_a$

denotes the each groove width, $\Lambda_i$ denotes the grooves width, and $h_i$ denotes the depth of grooves with the serial number of N. $h_d$ denotes the depth difference between the adjacent grooves, and $h_d>0$, $h_d=0$, $h_d<0$ represent the cases shown in a, b, and c, respectively.

The incident light is the radial polarization wave, impinging from the bottom of the substrate. The principle of excitation and transmission for CSP waves in metal slit are given in our previous work[22], and the tunable nanofocusing is formed on the basis of beam interference between the coupled SPP wave and diffraction wavelets generated from each ring of circular grating. In a homogeneous, constant-dielectric medium, the propagation vector is $k = k_0 \left( \sigma_r \vec{r} + \sigma_z \vec{z} \right)$, where $k_0 = 2\pi / \lambda_0$ and $\sigma_r^2 + \sigma_z^2 = \varepsilon_{metal}$. For our plasmonic structure, the E-field profile for glass layer, metal film, dielectric layer, and circular grating layer, can be given, respectively, as follows[23,24],

$$E_r = \begin{cases} E(z) e^{i\left[k_0\sigma_r r \pm \sigma_{air}(z-d_{glass}-d_2)\right]} & |z| \geq d_{glass} + d_2 \\ E(z) e^{i\left[k_0\sigma_r r \pm \sigma_{air}(z-d_2)\right]} & d_{glass} + d_2 \leq |z| \leq d_2 \\ E(z) e^{i\left[k_0\sigma_r r \pm \sigma_{air}(z-d_3)\right]} & d_2 \leq |z| \leq d_3 \\ E(z) e^{i\left[k_0\sigma_r r \pm \sigma_{air}(z-d_4)\right]} & d_3 \leq |z| \leq d_2 \end{cases} \quad (3)$$

where $d_{glass}$, $d_2$, $d_3$, and $d_4$ are the thickness of the glass substrate, metal film, dielectric layer, and circular grating layer, respectively. The $E_z$ and $H_\varphi$ can be deduced in the same way.

These electromagnetic field components can be modulatd by linear-variant depths and nonlinear-variant widths of circular grating in the output surface. The final intensity at the focal point is synthesized by iteration of each zone focusing and interference each other, and can be approximately expressed as[25]

$$I = \sum_i^N \alpha_i C_i \left( \sum_j^M C_j I_{j0,i} I_{SP,j} \frac{4r_j}{\lambda_{SP}} e^{-(r_j/l_{SP})} \right) e^{-\left(\sqrt{r_i^2 + h_i^2}/\lambda_0\right)} \quad (4)$$

where $\lambda_{SP}$ is the SPP wavelength as expressed in Eq.(5), $\lambda_0$ is the incident wavelength, $I_{SP,j}$ is the intensity of the SPP wave passing through the $j$th metal slit, $l_{SP}$ is the propagation length for the SPP wave, $\alpha_i$ is interference factor of $i$th circular grating, $I_{i0,j}$ is the intensity of diffractive wavelet at $j$th slit and $i$th grating, $h_i$ is height of $i$th grating, $r_i$ and $r_j$ is the radius of each zone of grating and slits, and $C_i$ and $C_j$ are the coupling efficiency of the grating and slits, $i$ and $j$ are the number of the zones, respectively. $C_i$ and $C_j$ are the complicated functions of the grating and slits geometry and will likely have different functional forms when the grating and slit width are much larger or much smaller than the incident wavelength. The interference phenomena originate from both the subwavelet $I_{j0}$ and surface plasmonic wave $I_{SP}$. The numerical simulation results are shown by using Finite element Method in section 3.

$$\lambda_{sp} = \lambda_0 \sqrt{\frac{\mathrm{Re}(\varepsilon_m) + \varepsilon_d}{\mathrm{Re}(\varepsilon_m) \cdot \varepsilon_d}} \quad (5)$$

## 3. NUMERICAL SIMULATIONS

The softwares of COMSOL Multiphysics 4.3a and Matlab are used for simulation. We assume that the device is illuminated by a plane wave in radial polarization. A blocking chip is also used to enhancing the focusing efficiency[22]. The substrate is glass ($n_1$=1.45). The circular grating layer is ZnO with $n_2$=2.16, and the dielectric layer is artificial quartz crystal ($\varepsilon_3 = 2.4$). The metal film is selected as titanium (Ti). CSP, which is excited at metal annular slits, is closely related to the permittivity of metal[18]. The image part of metal's permittivity is associated with the intrinsic losses of metal[3]. In our lens structure, it is benefit for the subwavelength annular slits exciting diffracted cylindrical waves to launch CSP when value of metal's intrinsic-loss is higher. The Drude model[26, 27] for titanium (Ti) is taken as

$$\varepsilon_4 = \varepsilon_{metal}' + i \cdot \varepsilon_{metal}'' = \left(1 - \omega_p^2 / (\omega^2 + i\omega\Gamma)\right) \quad (6)$$

where $\omega_p$ is the plasma frequency, $\Gamma$ the relaxation constant, and $\omega$ the angular frequency of the source. The imaginary part of Ti is increased as the wavelength increases, as shown in Fig.3.

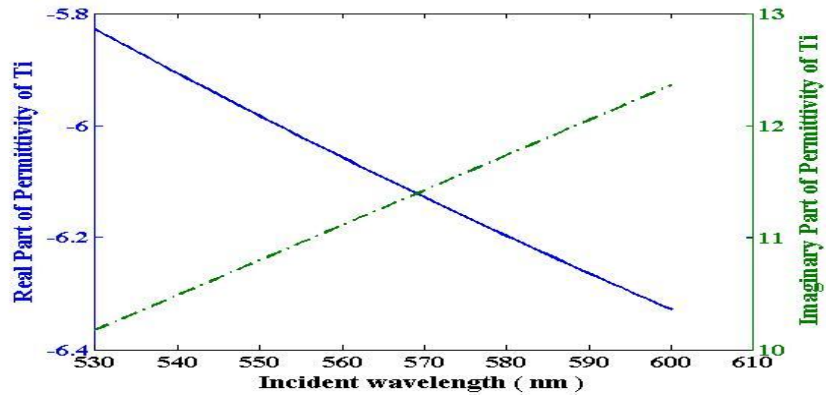

**Fig. 3**. The relationship for permittivity of Ti

The dispersion relationship in the glass-metallic-dielectric guide can be written in our previous work[5]. Since the polarization symmetry of a radially polarized illumination matches to the rotational symmetry cylinder structure, the entire incident beam is TM polarized with respect to the annular slit rings, enabling surface plasmon excitation from all directions. Part of the incident light is diffracted by the sharp edge of the slit, and then CSP waves are excited by the diffracted light for an extra wave vector in the direction along the film surface is obtained[28]. The glass-metallic-dielectric guide is an asymmetric structure, and there is the asymmetric surface plasmons (SPs) mode. Ref. [29] and Ref. [5] have shown that the group velocity of the asymmetric surface plasmons mode can be negative by engineering the dispersion curve. Because the group velocity[30] is defined as $v_g = d\omega/dk$, the slope negative at the large wave vector region when the incident wavelength is 587[nm], indicating that the group velocity is negative[29], as shown in Fig. 4. In contrast, the slope of the SPP dispersion curve is positive when the incident wavelength is 532[nm][18]. And this result is consistent with Ref.[5] and Ref. [29].

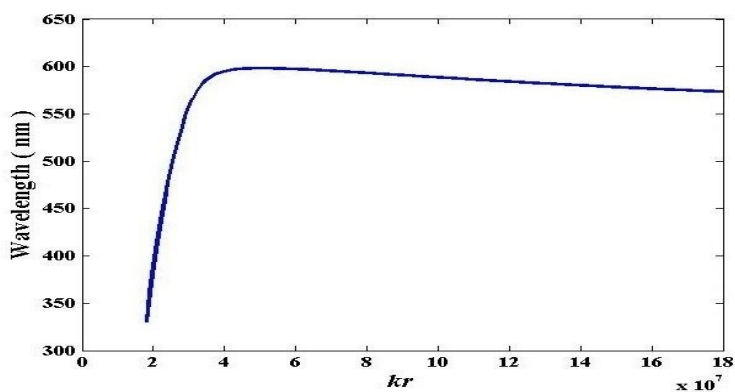

Fig. 4. Dispersion curve of surface plasmons for glass-metallic-dielectric structure

In order to obtain the optical energy which is transmitted into the our lens structure in this case, the reflection coefficient $R$ on the glass-metallic-dielectric guide can be calculated by considering the resonance of the multiply reflected field[1, 5]:

$$R = \left| \frac{r_{12} + r^1 e^{(2id_1k_1)}}{1 + r_{12}r^1 e^{(2i\omega d_1k_1)}} \right|^2 \qquad (7)$$

where $k_i = \sqrt{\varepsilon_i - \varepsilon_1 \sin^2\theta_i} \times \omega/c,\ i=1,2,3,4$, $r_{ij} = \left(\sqrt{\varepsilon_j}\cos\theta_i - \sqrt{\varepsilon_i}\cos\theta_j\right)/\left(\sqrt{\varepsilon_j}\cos\theta_i + \sqrt{\varepsilon_i}\cos\theta_j\right)$, $\theta_i$ is incident angle and $\theta_j$ is the refractive angle in $i$ layer, $r^1 = r_{23} + r^2 e^{(2id_2k_2)} / 1 + r_{23}r^2 e^{(2id_2k_2)}$, $r^2 = r_{34} + r_{41}e^{(2id_3k_3)} / 1 + r_{34}r_{41}e^{(2id_3k_3)}$. And Fig. 5 shows the reflection coefficient $R$ as a function of the incident angle. Only incident angle is 81.36°, the value of $R$ can be the minimum when the incident wavelength is 587[nm], which means that SPP has been excited and the most optical energy is coupled with CSP. So we choose 587[nm] as incident wavelength. In order to excite the CSP effectively, the width of slits (*w*) should be less than half of the incident wavelength. Then the Ti slab slits are 228[nm] in width, and the inner radii satisfy the principle of the destructive interference of SPPs as mentioned in Ref. [22].

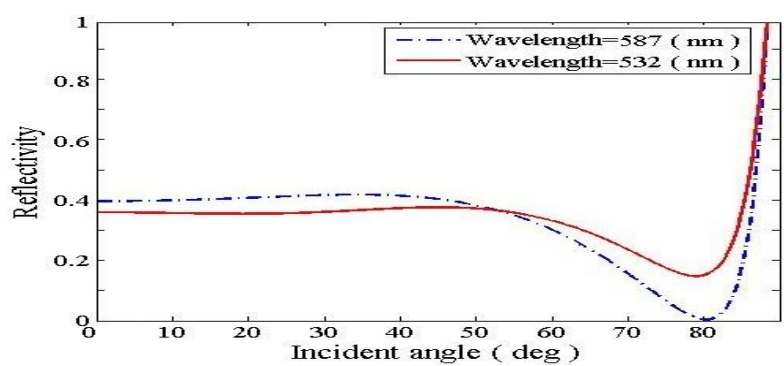

**Fig. 4**. The reflection coefficient *R* as a function of the incident angle

In following, we discuss the mechanism of focusing position shift. In order to adjust the focal length, a periodic concentric dielectric grooves is added topside, which can be treated as an annular surface grating limited by a convergent diaphragm, acting as a modulation device. Eq. (4) shows that CSP can be launch to far field by coupling into the circular grating. $C_i$ is the coupling efficiency of the grating and it is the complicated functions of the grating geometry. So the depth ($h_i$) and width ($\Lambda_i$) of circular grating play a key role in focusing position shift. The varying cross-section profile of circular grating can be expressed as Eq.(1)~(2), where $h_d$ denotes the depth and width difference between the adjacent grooves. It is called the positive change (PC) of the depths and widths of circular grating when $h_d > 0$, and it is called the negative change (NC) when $h_d < 0$. Fig. 5(a)~(c) illustrate the corresponding normalized intensity distributions for the case shown in Figs. 2(a)~(c). Fig. 6(a)~(c) illustrate the corresponding three - dimensional normalized intensity distributions.

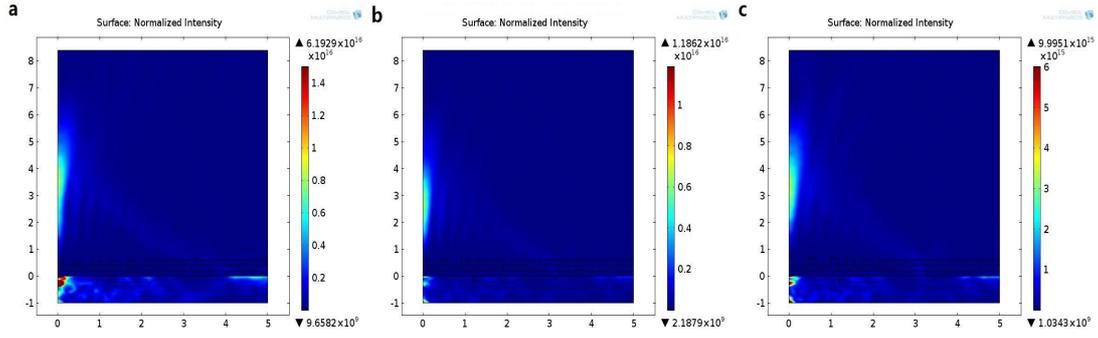

**Fig. 5**. The normalized intensity distributions: (a) for cases shown in Fig. 2(a) with $h_d$=20nm; (b) for cases shown in Fig. 2(b) with $h_d$=0nm; (c) for cases shown in Fig. 2(c) with $h_d$=-20nm;

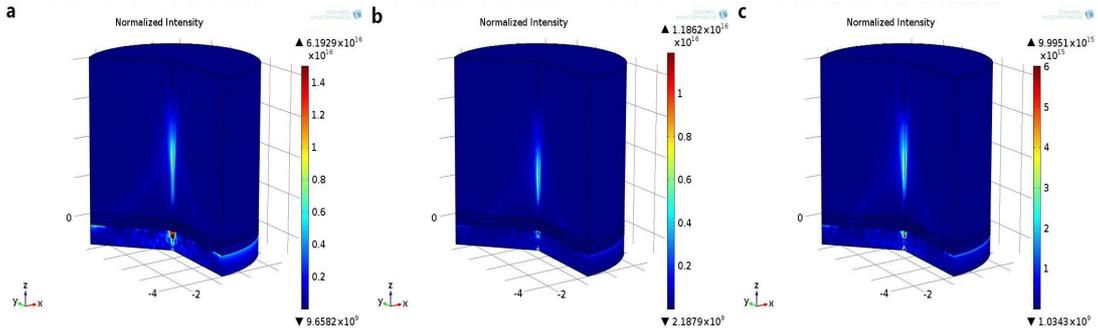

**Fig. 6**. The three - dimensional normalized intensity distributions: (a) for cases shown in Fig. 5(a) with $h_d$=20nm; (b) for cases shown in Fig. 5(b) with $h_d$=0nm; (c) for cases shown in Fig. 5(c) with $h_d$=-20nm;

The normalized intensity distribution results show that the energy emerging from the structure overlaps the axis within several microns, concentrating most of the energy in an extremely small region. For example, the $h_d = 20nm$ case shown in Fig. 5 (a) reveals the focal length of $3.52\,\mu m$ and the full width at half maximum (FWHM) of the focal width of $0.35\,\mu m$, i.e., focal spot smaller than a wavelength. If $h_d = 0nm$, the focal length will reduce to $2.67\,\mu m$ with the decreased focal width of $0.29\,\mu m$. On the contrary, $h_d = -20nm$ case shows a contrary performance with the increased focal length of $3.12\,\mu m$ and focal width of $0.34\,\mu m$. Fig. 7 shows the intensity distributions of the profile of the foci.

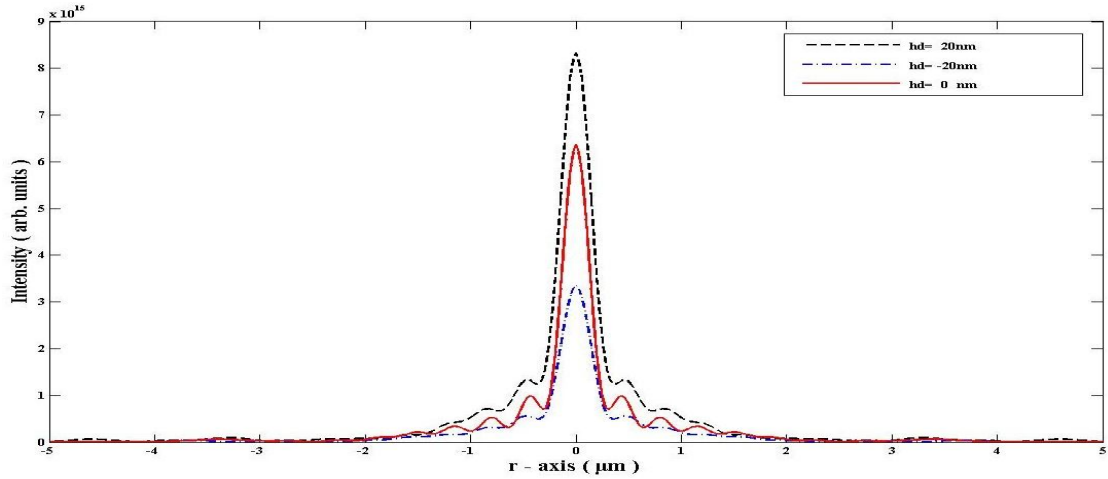

**Fig. 7**. The intensity distributions of the profile of the foci.

More details of the dependence of focal length and width on groove linear-variant depth and nonlinear-variant width trace profile are shown in Fig.8. The range of focusing points shift is about 1.94 $\mu m$ ( $3.3\lambda$ ) to 3.52 $\mu m$ ( $6\lambda$ ), which is much more than results of other lens[13-15, 20-22]. This adjusted range is much more than results as mentioned in Ref. [14] and Ref. [22]. It is worth noting that the focal length and focal width change of ups and downs by increasing $h_d$ from -30 to 30 $nm$ at a step of 10 $nm$. These results are different to Ref. [14] and Ref. [22]. The reason is that the widths of circular grating are nonlinear variation, and depths are linear variation. These superimposed variation results can cause that the relative phase at the circular groove exit change of ups and downs, as shown in Fig. 9.

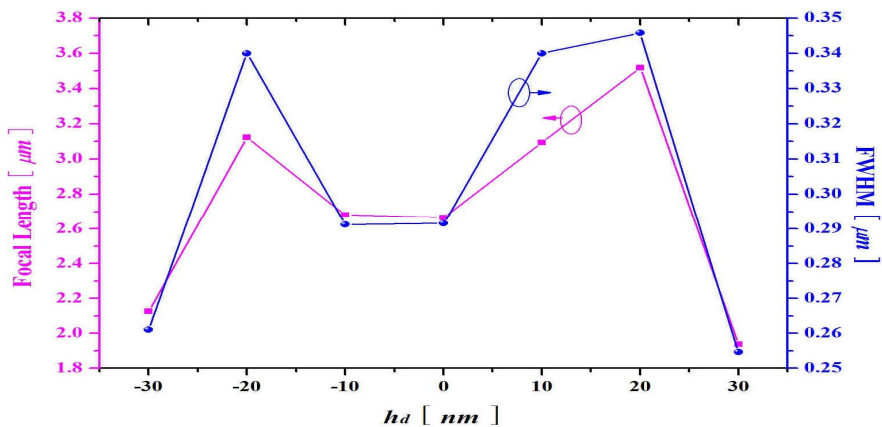

**Fig. 8**. The focal length and focal width vary with $h_d$. $h_d$ increases from -30 to 30 nm with a step of 10 nm, and the other parameters are the same as these used in Fig. 5.

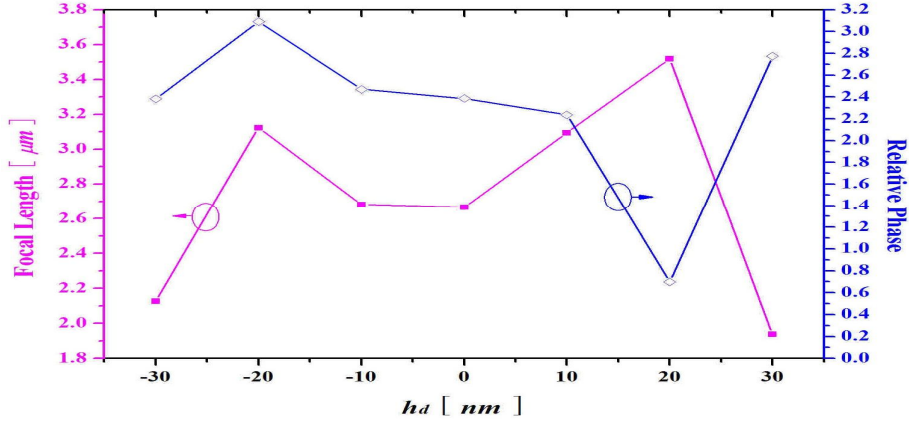

**Fig. 9**. The focal length and relative phase vary with $h_d$. $h_d$ increases from -30 to 30 nm with a step of 10 nm, and the other parameters are the same as these used in Fig. 5.

From Fig. 9, phase changing rules during PC and NC are different. In PC case, the rule of focusing points shift is the same as phase changing rule. On the contrary, the rule of focusing points shift has an inverse relationship to phase changing rule in NC case. Because the each depth and width of circular grating decreases or increases in PC or NC, respectively, the contributions to the phase are different. It is a unique phenomenon for the circular grating, which width is nonlinear variation, and depth is linear variation.

## 4. CONCLUSION

In this paper, we design a tunable nanofocusing lens capable of adjusting subwavelength focusing in far field under the nonlinear-variant widths and linear-variant depths of circular grating. Numerically simulations show that this adjusted range is much more than other plasmonic lens. We have revealed the relation between the linear-variant depths and nonlinear-variant widths of grooves and the relative phase of CSP in detail. Our far-field tunable nanofocusing scheme can supply for the requirement of practical applications such as nanolithography, superresolution optical microscopic imaging, optical trapping, and sensing.

## ACKNOWLEDGMENT

This work was supported by National Natural Science Foundation of China (No.61205204), "Spring sunshine"plan (No.Z2011029), the Fundamental Research Funds for the Central Universities (lzujbky-2012-42), and the Open Research Fund of State Key Laboratory of